\documentclass{article}
\usepackage{amsfonts}
\usepackage{mathrsfs}
\usepackage{amsthm}
\usepackage{amssymb}
\usepackage{amsmath}
\usepackage{enumerate}

\theoremstyle{plain}
\newtheorem{theorem}{Theorem}[section]
\newtheorem{proposition}[theorem]{Proposition}

\theoremstyle{definition}
\newtheorem{definition}{Definition}[section]

\theoremstyle{remark}
\newtheorem{remark}{\textbf{Remark}}[section]

\theoremstyle{example}

\numberwithin{equation}{section}

\title{Spectral Property of Magnetic Quantum Walk on Hypercube}
\author{Ce Wang \footnote{Author to whom correspondence should be addressed: cewangwhu@163.com}
\footnote{Current address: Shanghai Institute for Mathematics and Interdisciplinary Sciences,
Shanghai 200433, People's Republic of China}\\
Yau Mathematical Sciences Center, Tsinghua University\\
Beijing 100084, People's Republic of China}
\date{}
\begin{document}
\maketitle

\noindent\textbf{Abstract.}\ \
In this paper, we introduce and investigate a model of magnetic quantum walk on a general hypercube.
We first construct a set of unitary involutions associated with a magnetic potential $\nu$ by using
quantum Bernoulli noises. And then, with these unitary involutions as the magnetic shift operators,
we define the evolution operator $\mathsf{W}^{(\nu)}$ for the model, where $\nu$ is the magnetic potential.
We examine the point-spectrum and approximate-spectrum of the evolution operator $\mathsf{W}^{(\nu)}$
and obtain their representations in terms of the coin operator system of the model.
We show that the point-spectrum and approximate-spectrum of $\mathsf{W}^{(\nu)}$ are completely independent of
the magnetic potential $\nu$ although $\mathsf{W}^{(\nu)}$ itself is dependent of the magnetic potential $\nu$.
Our work might suggest that a quantum walk perturbed by a magnetic field can have spectral stability with respect
to the magnetic potential.
\vskip 2mm

\noindent\textbf{Keywords.}\ \ Magnetic quantum walk; Hypercube; Spectral property; Quantum Bernoulli noises
\vskip 2mm

\noindent\textbf{Mathematics Subject Classification.}\ \ 81S25; 60H40; 81Q99.

\section{Introduction}\label{sec-1}

The past two and half decades have seen great attention paid to quantum walks, which were originally introduced
by Aharonov \cite{Aharonov} as quantum analogs of classical random walks in probability theory
and now play an important role in many research fields such as quantum computing, quantum simulation and quantum probability
(see, e.g. \cite{Grover, Kempe, Venegas} and references therein).
There are two basic categories of quantum walks: discrete-time quantum walks and continuous-time quantum walks.
In this paper, we will only deal with discrete-time ones, which will be simply called quantum walks or even walks hereafter.

Mathematically, a quantum walk involves two Hilbert spaces: one is known as the position space,
the other is referred to the coin space. The position space is usually an $l^2$-space associated with a graph,
which describes the position information of the quantum walker, while the coin space is often a finite-dimensional space,
which reflects the quantum walker's internal degrees of freedom. The tensor of the position and coin spaces then
acts as the state space for the walk. For example, as a quantum walk on the $1$-dimensional integer lattice $\mathbb{Z}$
(as is known, $\mathbb{Z}$ is a regular infinite graph), the well-known Hadamard walk uses $l^2(\mathbb{Z})$ and $\mathbb{C}^2$
as its position and coin spaces respectively, hence its state space is $l^2(\mathbb{Z})\otimes \mathbb{C}^2$.

Considerable researches have been made concerning quantum walks on the $d$-dimensional integer lattice $\mathbb{Z}^d$,
especially on $\mathbb{Z}$ (see, e.g. \cite{Konno, Portugal, Venegas} and references therein).
In recent years, there has been much interest in quantum walks on hypercubes,
which belong to the category of regular finite graphs and play an important role in computer science.
Moore and Russell \cite{Moore} discussed the instantaneous mixing time of quantum walks on a general hypercube,
while Kempe \cite{Kempe-2005} considered the hitting time of such walks.
Alagi\'{c} and Russell \cite{Alagic} analyzed decoherence in quantum walks
on hypercubes and obtained interesting results. Recently, the author \cite{W-JMP-2022} introduced a model of quantum walk
on a general hypercube by using quantum Bernoulli noises, and investigated its behavior from a perspective of probability distribution.
It was shown in \cite{W-JMP-2022} that the walk introduced therein produces the uniform measure as its stationary measure on the hypercube
provided the initial state satisfies some mild conditions, among others.

A magnetic quantum walk is a quantum walk coupling to (or perturbed by) a magnetic field,
which, as pointed out in \cite{Cedzich-2019}, plays an active role in describing the motion of a particle in a magnetic field.
In 2015, Yalcinkaya and Gedik \cite{Yalcinkaya-2015} studied the dynamics of a magnetic quantum walk on $\mathbb{Z}^2$.
In 2020, Cedzich et al. \cite{Cedzich-2020} considered a similar magnetic quantum walk and found that the spectrum
is a zero-measure Cantor set provided the magnetic flux satisfies certain conditions.
%Recently, Yang \cite{Yang-2022} has proven Anderson localization for all Diophantine frequencies and
%all nonresonant phases in a magnetic quantum walk model on $\mathbb{Z}^2$.

In this paper, we consider a magnetized version of the quantum walk introduced in \cite{W-JMP-2022}.
More precisely, we will introduce a model of magnetic quantum walk on a general hypercube by using quantum Bernoulli noises,
and then investigate its behavior from a viewpoint of spectral theory. Our main work is as follows.
\begin{itemize}
  \item We construct a set of unitary involutions associated with a magnetic potential $\nu$ by using quantum Bernoulli noises,
        and then, with these unitary involutions as the magnetic shift operators,
        we define a model of magnetic quantum walk on the corresponding hypercube,
        whose evolution operator is written $\mathsf{W}^{(\nu)}$ (see Section~\ref{sec-3} below).
  \item After proving some technical results, we examine the point-spectrum and approximate-spectrum
        of the evolution operator $\mathsf{W}^{(\nu)}$ and obtain their representations in terms of the coin operator system $\mathfrak{C}$
        of the model (see Theorems~\ref{thr-point-spectrum} and \ref{thr-approximate-spectrum} bellow).
  \item We find that the point-spectrum and approximate-spectrum of $\mathsf{W}^{(\nu)}$ are completely independent of
        the magnetic potential $\nu$ although $\mathsf{W}^{(\nu)}$ itself is dependent of the magnetic potential $\nu$
        (see Theorems~\ref{thr-spectral-stability} below), which is of much physical interest.
\end{itemize}
Our work might suggest that a quantum walk perturbed by a magnetic field can have spectral stability with respect to the magnetic potential.

\section{Preliminaries}\label{sec-2}

This section recall some necessary notions and facts about Bernoulli functionals and quantum Bernoulli noises. For more details,
we refer to \cite{WCL-JMP-2010} or \cite{W-JMP-2022}.

Let $\mathfrak{h}$ be the $L^2$-space of complex-valued Bernoulli functionals, whose inner product and norm are denoted by
$\langle\cdot,\cdot\rangle_{\mathfrak{h}}$ and $\|\cdot\|_{\mathfrak{h}}$, respectively. It is known that $\mathfrak{h}$ has
an orthonormal basis of the form $\{Z_{\sigma}\mid \sigma \in \Gamma\}$, where $\Gamma$ stands for the finite power set
of the nonnegative integer set $\mathbb{N}$. Hence $\mathfrak{h}$ is separable and infinitely-dimensional.

For each $k\in \mathbb{N}$, the annihilation operator $\partial_k$ associated with $k$ is the one
on $\mathfrak{h}$ determined by
\begin{equation}\label{eq-annihilation}
  \partial_kZ_{\sigma}= \mathbf{1}_{\sigma}(k)Z_{\sigma\setminus k},\quad \sigma\in \Gamma,
\end{equation}
where $\mathbf{1}_{\sigma}(\cdot)$ is the indicator of the set $\sigma$ and $\sigma\setminus k$ is the shorthand for
the difference set $\sigma\setminus \{k\}$. In the literature, the operator family $\{\partial_k,\, \partial_k^*\mid k\in \mathbb{N}\}$
is known as quantum Bernoulli noises (QBN), where $\partial_k^*$ is the adjoint of $\partial_k$ and is called the creation operator
associated with $k$, which satisfies that
\begin{equation}\label{eq-creation}
  \partial_k^*Z_{\sigma}= (1-\mathbf{1}_{\sigma}(k))Z_{\sigma\cup k},\quad \sigma\in \Gamma.
\end{equation}
One typical property of QBN is that it satisfies the canonical anti-commutation relation (CAR) in equal time
\begin{equation}\label{eq-2-3}
  \partial_k^*\partial_k + \partial_k\partial_k^* = I,
\end{equation}
where $I$ is the identity operator on $\mathfrak{h}$ and $k$ ranges over $\mathbb{N}$.

For a nonnegative integer $n \geq 0$, let $\Gamma_n$ denote the power set of the set $\mathbb{N}_n:=\{0,1,2,\cdots,n\}$ and
\begin{equation}\label{eq-2-4}
\mathfrak{h}_n = \mathrm{Span}\{Z_{\sigma} \mid \sigma\in \Gamma_n\},
\end{equation}
namely $\mathfrak{h}_n$ is the linear subspace of $\mathfrak{h}$ spanned by the orthonormal system $\{Z_{\sigma} \mid \sigma\in \Gamma_n\}$.
Then, with the linear operation in $\mathfrak{h}$ and the inner product $\langle\cdot,\cdot\rangle_{\mathfrak{h}}$, $\mathfrak{h}_n$
becomes a $2^{n+1}$-dimensional complex Hilbert space. Clearly the vector system
$\mathfrak{Z}_n:=\{Z_{\sigma} \mid \sigma\in \Gamma_n\}$ is an orthonormal basis (ONB)
for $\mathfrak{h}_n$, which we call the canonical ONB for $\mathfrak{h}_n$ in the sequel.

\begin{remark}\label{rem-2-1}
Let $n \geq 0$ be a nonnegative integer. Then, for each $k\in \mathbb{N}_n$, both $\partial_k$ and $\partial_k^*$
leave $\mathfrak{h}_n$ invariant, and moreover $(\partial_k|_{\mathfrak{h}_n})^*= \partial_k^*|_{\mathfrak{h}_n}$,
where $\partial_k|_{\mathfrak{h}_n}$ and $\partial_k^*|_{\mathfrak{h}_n}$ mean the restrictions of $\partial_k$ and $\partial_k^*$
to $\mathfrak{h}_n$ respectively and $(\partial_k|_{\mathfrak{h}_n})^*$ is the adjoint of $\partial_k|_{\mathfrak{h}_n}$ on $\mathfrak{h}_n$.
\end{remark}

According to this remark, for each $k\in \mathbb{N}_n$, one can naturally view $\partial_k$ and $\partial_k^*$ as
operators on $\mathfrak{h}_n$. In that case, $\partial_k$ and $\partial_k^*$ are mutually adjoint,
and moreover it holds that
\begin{equation}\label{eq-CAR-1}
\partial_j\partial_k=\partial_k\partial_j,\quad \partial_j^*\partial_k^*=\partial_k^*\partial_j^*,\quad \partial_j^*\partial_k=\partial_k\partial_j^*\qquad (j,\, k \in \mathbb{N}_n,\, j\ne k)
\end{equation}
and
\begin{equation}\label{eq-CAR-2}
\partial_j\partial_j=\partial_j^*\partial_j^*=0,\quad
\partial_j^*\partial_j + \partial_j\partial_j^* = I_{\mathfrak{h}_n},
\qquad (j\in \mathbb{N}_n),
\end{equation}
where $I_{\mathfrak{h}_n}$ denotes the identity operator on $\mathfrak{h}_n$.

\section{Magnetic quantum walk}\label{sec-3}

In this section, we describe our model of magnetic quantum walk and examine its basic properties. Unless otherwise specified,
$n\geq 0$ is assumed to be a fixed nonnegative integer hereafter.
Additionally, we are given a complex separable Hilbert space $\mathcal{K}$ with $\dim \mathcal{K}\geq n+1$,
which has an orthonormal basis $\{e_j \mid 0\leq j \leq d_{\mathcal{K}}-1\}$, where $d_{\mathcal{K}}=\dim \mathcal{K}$.
It may happen that $\dim \mathcal{K}=\infty$. In that case we follow the convention that $\infty -1 =\infty$.

Consider $\Gamma_n$, the power set of the set $\mathbb{N}_n=\{0, 1, \cdots, n\}$. Two elements $\sigma$ and $\tau$ in $\Gamma_n$
are said to be adjacent if $\#(\sigma\triangle \tau)=1$. In that case, we writes $\sigma\sim \tau$.
Note that, as a regular graph, $(\Gamma_n,\sim)$ is isomorphic to the standard $(n+1)$-dimensional hypercube \cite{W-JMP-2022}.

\begin{definition}
By a magnetic potential on $\Gamma_n$ we mean a function $\nu\colon \Gamma_n\times \Gamma_n\rightarrow [-\pi,\pi]$ that satisfies
$\nu(\sigma,\tau) = -\nu(\tau,\sigma)$, $\forall\, \sigma$, $\tau\in \Gamma_n$.
\end{definition}

Apparently, the function $\mathbf{0}\colon \Gamma_n\times \Gamma_n\rightarrow [-\pi,\pi]$ given by $\mathbf{0}(\sigma,\tau)=0$,
$(\sigma, \tau)\in \Gamma_n\times \Gamma_n$, is a magnetic potential on $\Gamma_n$, which we call the null magnetic potential on $\Gamma_n$.

\textbf{In the rest of the paper, we assume that $\nu$ is a given magnetic potential on $\Gamma_n$}.
We use the following simplified notation
\begin{equation}\label{eq}
  \nu_j = \nu(\{j\}, \mathbb{N}_n\setminus \{j\}),\quad j\in \mathbb{N}_n,
\end{equation}
where $\mathbb{N}_n\setminus \{j\}$ is the difference of sets $\mathbb{N}_n$ and $\{j\}$.
Physically, the set $\{\nu_j\mid j\in \mathbb{N}_n\}$ still contains much information of interest about the magnetic potential $\nu$.

As is seen in Section~\ref{sec-2}, $\mathfrak{h}_n$ is a Hilbert space of dimension $2^{n+1}$. The next theorem shows that
the magnetic potential $\nu$ can give rise to some unitary operators of physical interest.

\begin{theorem}\label{thr-3-1}
Given $j\in \mathbb{N}_n$, the operator $\Xi_j^{(\nu)}:= e^{-\mathrm{i}\nu_j}\partial_j^* + e^{\mathrm{i}\nu_j}\partial_j$
is a  unitary involution on $\mathfrak{h}_n$. Moreover, two vertices $\sigma$, $\tau\in \Gamma_n$ are adjacent if and only if
there exists a unique $j\in \mathbb{N}_n$ such that
\begin{equation}\label{eq-Mshift}
   \Xi_j^{(\nu)}Z_{\sigma}
     = \big[(1-\mathbf{1}_{\sigma}(j))e^{-\mathrm{i}\nu_j} + \mathbf{1}_{\sigma}(j)e^{\mathrm{i}\nu_j}\big]Z_{\tau}.
\end{equation}
\end{theorem}

\begin{proof}
For $j\in \mathbb{N}_n$, it follows easily from (\ref{eq-CAR-2}) that $\Xi_j^{(\nu)}$ is a self-adjoint unitary operator,
i.e. a unitary involution. Next, let us prove the second part of the theorem.

Suppose $\sigma$, $\tau\in \Gamma_n$ are adjacent, namely $\sigma\sim \tau$. By the definition, there is a unique $j\in \mathbb{N}_n$
such that $\sigma\triangle\tau=\{j\}$, equivalently $\tau = \sigma\triangle j$. Thus, by the definition of $\Xi_j^{(\nu)}$ and
properties (\ref{eq-annihilation}) and (\ref{eq-creation}), we have
\begin{equation}\label{eq-Mshif-action}
  \Xi_j^{(\nu)}Z_{\sigma}
     = (1-\mathbf{1}_{\sigma}(j))e^{-\mathrm{i}\nu_j}Z_{\sigma\cup j}
       + \mathbf{1}_{\sigma}(j)e^{\mathrm{i}\nu_j}Z_{\sigma\setminus j}
     = \big[(1-\mathbf{1}_{\sigma}(j))e^{-\mathrm{i}\nu_j}
       + \mathbf{1}_{\sigma}(j)e^{\mathrm{i}\nu_j}\big]Z_{\sigma\triangle j},
\end{equation}
which together with $\tau = \sigma\triangle j$ yields (\ref{eq-Mshift}). Now, suppose there exists a unique $j\in \mathbb{N}_{n}$ such that
(\ref{eq-Mshift}) holds. Then, by using (\ref{eq-Mshif-action}), we immediately find
\begin{equation*}
  \big[(1-\mathbf{1}_{\sigma}(j))e^{-\mathrm{i}\nu_j} + \mathbf{1}_{\sigma}(j)e^{\mathrm{i}\nu_j}\big]Z_{\tau}
  = \big[(1-\mathbf{1}_{\sigma}(j))e^{-\mathrm{i}\nu_j} + \mathbf{1}_{\sigma}(j)e^{\mathrm{i}\nu_j}\big]Z_{\sigma\triangle j},
\end{equation*}
which means that $Z_{\tau}=Z_{\sigma\triangle j}$, hence $\tau = \sigma\triangle k$, equivalently $\sigma\triangle\tau =\{j\}$.
Thus $\sigma\sim \tau$.
\end{proof}

Recall that $\mathfrak{Z}_n=\{Z_{\sigma} \mid \sigma\in \Gamma_n\}$, the canonical ONB for $\mathfrak{h}_n$,
is exactly indexed by the vertex set $\Gamma_n$ of the graph $(\Gamma_n,\sim)$. On the other hand, given $j\in \mathbb{N}_n$
and $\sigma\in \Gamma_n$, one has
\begin{equation*}
  \Xi_j^{(\nu)}Z_{\sigma}
     = \big[(1-\mathbf{1}_{\sigma}(j))e^{-\mathrm{i}\nu_j}
       + \mathbf{1}_{\sigma}(j)e^{\mathrm{i}\nu_j}\big]Z_{\sigma\triangle j}.
\end{equation*}
This, together with the fact that $\sigma$ and $\sigma\triangle j$ are adjacent as vertices of the graph $(\Gamma_n,\sim)$,
suggests that the unitary involution $\Xi_j^{(\nu)}$ can actually be interpreted as a magnetic shift operator on the graph $(\Gamma_n, \sim)$.

\begin{proposition}\label{prop-shift-commutative}
Any two operators in the system $\big\{\Xi_j^{(\nu)} \mid j\in \mathbb{N}_n\big\}$ are commutative. Moreover,
for $\gamma\in \Gamma_n$, it holds that
\begin{equation}\label{eq-MSO-product-action}
   \Big(\prod_{j\in\gamma}\Xi_j^{(\nu)}\Big)Z_{\emptyset} = e^{-\mathrm{i}\sum_{j\in\gamma}\nu_j}Z_{\gamma},\quad
  \Big(\prod_{j\in\gamma}\Xi_j^{(\nu)}\Big)Z_{\gamma} = e^{\mathrm{i}\sum_{j\in\gamma}\nu_j}Z_{\emptyset},
\end{equation}
where $Z_{\emptyset}\in \mathfrak{Z}_n$ is the basis vector indexed by the empty subset $\emptyset$ of $\mathbb{N}_n$, and
$\prod_{j\in\gamma}\Xi_j^{(\nu)}=I_{\mathfrak{h}_n}$, $\sum_{j\in\gamma}\nu_j=0$ when $\gamma=\emptyset$.
\end{proposition}

\begin{proof}
The commuting property of the system $\big\{\Xi_j^{(\nu)} \mid j\in \mathbb{N}_n\big\}$ is an immediate consequence
of (\ref{eq-CAR-1}) and (\ref{eq-CAR-2}).
For $\gamma\in \Gamma_n$, by the definition of $\Xi_j^{(\nu)}$ and properties (\ref{eq-annihilation}) and (\ref{eq-creation}),
we can get
\begin{equation*}
   \Big(\prod_{j\in\gamma}\Xi_j^{(\nu)}\Big)Z_{\emptyset} = e^{-\mathrm{i}\sum_{j\in\gamma}\nu_j}Z_{\gamma},
\end{equation*}
which implies the second formula in (\ref{eq-MSO-product-action}).
\end{proof}

\begin{definition}\cite{W-JMP-2022}\label{def-coin-operator}
An operator system $\mathfrak{C}=\{C_j \mid 0\leq j \leq n\}$ on $\mathcal{K}$ is said to be a coin operator system
if its sum $\sum_{j=0}^n C_j$ is a unitary operator on $\mathcal{K}$ and
\begin{equation}\label{eq-coin-operator-property}
C^*_jC_k= C_j C_k^*=0,\quad  j\neq k,\, 0\leq j,\, k \leq n,
\end{equation}
where $C_j^*$ denotes the adjoint of $C_j$.
\end{definition}

\textbf{In the reminder, we also assume that $\mathfrak{C}=\{C_j \mid 0\leq j \leq n\}$ is a fixed coin operator system on $\mathcal{K}$}.
For $\sigma\in \Gamma_n$, we further define
\begin{equation}\label{eq-weighted-coin-sum}
  U_{\sigma}= \sum_{j=0}^n \mathcal{E}_{\sigma}(j)C_j,
\end{equation}
and call it the algebraic sum of the system $\mathfrak{C}$ corresponding to $\sigma$,
where $\mathcal{E}_{\sigma}(j)= 2\times \mathbf{1}_{\sigma}(j)-1$.
As was shown in \cite{W-JMP-2022}, $U_{\sigma}$ is a unitary operator on $\mathcal{K}$ for all $\sigma\in \Gamma_n$.

\begin{theorem}\label{thr-evolution-operator}
$\mathsf{W}^{(\nu)}:= \sum_{j=0}^n \Xi_j^{(\nu)}\otimes C_j$ is a unitary operator on the tensor space $\mathfrak{h}_n\otimes \mathcal{K}$.
\end{theorem}

\begin{proof}
For $j\in \mathbb{N}_n$, since $\Xi_j^{(\nu)}$ is a unitary involution (by Theorem~\ref{thr-3-1}),
$\big(\Xi_j^{(\nu)}\big)^{*} = \Xi_j^{(\nu)}$ and $\big(\Xi_j^{(\nu)}\big)^2=I_{\mathfrak{h}_{n}}$.
Thus, using properties of the coin operator system $\mathfrak{C}$ gives
\begin{equation*}
\begin{aligned}
  \big(\mathsf{W}^{(\nu)}\big)^*\mathsf{W}^{(\nu)}
  &=\Big(\sum_{j=0}^n\Xi_j^{(\nu)}\otimes C_j^*\Big)\Big(\sum_{j=0}^n\Xi_j^{(\nu)}\otimes C_j\Big)\\
  &=\sum_{j=0}^n\big(\Xi_j^{(\nu)}\big)^{2} \otimes C_j^{*}C_j
   =\sum_{j=0}^nI_{\mathfrak{h}_n}\otimes C_j^*C_j
   =I_{\mathfrak{h}_n}\otimes I_{\mathcal{K}}
   =I,
\end{aligned}
\end{equation*}
where $I$ denotes the identity operator on $\mathfrak{h}_n\otimes \mathcal{K}$.
Similarly, we have $\mathsf{W}^{(\nu)}\big(\mathsf{W}^{(\nu)}\big)^* =I$.
\end{proof}

We are now ready to give the precise description of our magnetic quantum walk. As mentioned previously, $\mathbb{N}$ denotes the set of
all nonnegative integers.

\begin{definition}\label{def-walk}
The magnetic quantum walk we deal with is the quantum walk on the graph $(\Gamma_n,\sim)$ that takes
$\mathsf{W}^{(\nu)}$ as its evolution operator:
\begin{equation}
     \Phi_{t+1} = \mathsf{W}^{(\nu)}\Phi_t,\qquad t \in \mathbb{N},
\end{equation}
where $\Phi_t$ stands for the state of the walk at time $t$.
\end{definition}

By convention, we simply use $\mathsf{W}^{(\nu)}$ to stand for the magnetic quantum walk described above.
In that case, we just say the walk $\mathsf{W}^{(\nu)}$. For convenience, we also refer to $\mathfrak{C}=\{C_j \mid 0\leq j \leq n\}$
as the coin operator system of the walk $\mathsf{W}^{(\nu)}$. Note the coin operator system $\mathfrak{C}$ is independent of
the magnetic potential $\nu$ although the walk $\mathsf{W}^{(\nu)}$ itself is dependent of $\nu$.

As is seen, the walk $\mathsf{W}^{(\nu)}$ takes the tensor space $\mathfrak{h}_n\otimes \mathcal{K}$ as its state space
and its states are represented by unit vectors in $\mathfrak{h}_n\otimes \mathcal{K}$.
In particular, $\mathfrak{h}_n$ and $\mathcal{K}$ serve as the position and coin spaces for the walk $\mathsf{W}^{(\nu)}$, respectively.

\begin{remark}
The magnetic quantum walk introduced in Definition~\ref{def-walk} is actually a magnetized version of that introduced in \cite{W-JMP-2022}.
In fact, with $\nu \equiv 0$ the walk $\mathsf{W}^{(\nu)}$ becomes the walk introduced in \cite{W-JMP-2022}.
\end{remark}

\section{Spectral property}\label{sec-4}

This section investigates the walk $\mathsf{W}^{(\nu)}$ from a viewpoint of spectral theory.
More specifically, we will examine spectral property of $\mathsf{W}^{(\nu)}$ as a unitary operator on $\mathfrak{h}_n\otimes \mathcal{K}$.
We keep using the notation and assumptions built in previous sections.

We first prove some technical propositions and theorems. Consider the unitary involutions
$\Xi_j^{(\nu)}$, $j\in \mathbb{N}_n$, which are defined in Theorem~\ref{thr-3-1}.
With these unitary involutions, we construct a new system of operators as follows:
\begin{equation}\label{eq-hat-product-operator}
 \widehat{\Xi}_{\sigma}^{(\nu)}=\prod_{j=0}^n\Big(I_{\mathfrak{h}_n}+\mathcal{E}_{\sigma}(j)\Xi_j^{(\nu)}\Big),\quad \sigma\in\Gamma_n.
\end{equation}
Here $\nu$ is the magnetic potential on $\Gamma_n$.
By (\ref{eq-CAR-1}) and (\ref{eq-CAR-2}), one immediately knows that each $\Xi_{\sigma}^{(\nu)}$ is a self-adjoint operator on $\mathfrak{h}_n$.

\begin{proposition}\label{pro-operator-eigenvector}
Given $\sigma\in\Gamma_n$ and $j\in\mathbb{N}_n$, it holds true that
\begin{equation}
  \Xi_j^{(\nu)}\widehat{\Xi}_{\sigma}^{(\nu)} = \mathcal{E}_{\sigma}(j)\widehat{\Xi}_{\sigma}^{(\nu)}.
\end{equation}
\end{proposition}

\begin{proof}
Let $\sigma\in\Gamma_{n}$ and $j\in\mathbb{N}_{n}$ be given. Then, with the fact that $\Xi_j^{(\nu)}$ is a unitary involution, we have
\begin{equation*}
  \Xi_j^{(\nu)}\big(I_{\mathfrak{h}_n}+\mathcal{E}_{\sigma}(j)\Xi_j^{(\nu)}\big)
  = \Xi_j^{(\nu)}+\mathcal{E}_{\sigma}(j)I_{\mathfrak{h}_n}
  =\mathcal{E}_{\sigma}(j)\big(I_{\mathfrak{h}_n}+\mathcal{E}_{\sigma}(j)\Xi_j^{(\nu)}\big).
\end{equation*}
According to (\ref{eq-CAR-1}) and (\ref{eq-CAR-2}), any two operators in
$\big\{I_{\mathfrak{h}_n}+\mathcal{E}_{\sigma}(j)\Xi_j^{(\nu)}\mid  j\in \mathbb{N}_n\}$ are commutative. Thus

\begin{equation*}
\begin{aligned}
  \Xi_j^{(\nu)}\widehat{\Xi}_{\sigma}^{(\nu)}
  & = \Xi_j^{(\nu)}\big(I_{\mathfrak{h}_n}+\mathcal{E}_{\sigma}(j)\Xi_j^{(\nu)}\big)
      \prod_{k\ne j}\big(I_{\mathfrak{h}_n}+\mathcal{E}_{\sigma}(k)\Xi_k^{(\nu)}\big)\\
  & = \mathcal{E}_{\sigma}(j)\big(I_{\mathfrak{h}_n}+\mathcal{E}_{\sigma}(j)\Xi_j^{(\nu)}\big)
       \prod_{k\ne j}\big(I_{\mathfrak{h}_n}+\mathcal{E}_{\sigma}(k)\Xi_k^{(\nu)}\big)\\
  & = \mathcal{E}_{\sigma}(j)\prod_{k}\big(I_{\mathfrak{h}_n}+\mathcal{E}_{\sigma}(k)\Xi_k^{(\nu)}\big)\\
  &=\mathcal{E}_{\sigma}(j)\widehat{\Xi}_{\sigma}^{(\nu)},
\end{aligned}
\end{equation*}
where $\prod_{k\ne j}$ means $\prod_{k=0,k\ne j}^n$, and $\prod_{k}$ means $\prod_{k=0}^n$.
\end{proof}

\begin{theorem}\label{thr-hat-ONB}
There exists an ONB\ \ $\widehat{\mathfrak{Z}}_n^{(\nu)}=\big\{\widehat{Z}_{\sigma}^{(\nu)} \mid \sigma\in \Gamma_n\big\}$
for $\mathfrak{h}_n$ such that
\begin{equation}\label{eq-hat-ONB}
  \Xi_j^{(\nu)}\widehat{Z}_{\sigma}^{(\nu)}= \mathcal{E}_{\sigma}(j)\widehat{Z}_{\sigma}^{(\nu)}
\end{equation}
holds for all $j\in \mathbb{N}_n$ and $\sigma\in \Gamma_n$.
\end{theorem}

\begin{proof}
Consider the canonical ONB for $\mathfrak{h}_n$, which is of the form $\mathfrak{Z}_n=\{Z_{\sigma}\mid \sigma\in\Gamma_n\}$.
Note that $\emptyset \in \Gamma_n$, which implies that $Z_{\emptyset}\in \mathfrak{Z}_n$. Now,
using operators defined in (\ref{eq-hat-product-operator}) and the basis vector $Z_{\emptyset}$, we can construct
a vector system $\widehat{\mathfrak{Z}}_n^{(\nu)}=\big\{\widehat{Z}_{\sigma}^{(\nu)} \mid \sigma\in \Gamma_n\big\}$
in $\mathfrak{h}_n$ as follows:
\begin{equation}\label{eq-hat-ONB-vector}
\widehat{Z}_{\sigma}^{(\nu)}=\frac{1}{\sqrt{2^{n+1}}}\widehat{\Xi}_{\sigma}^{(\nu)}Z_{\emptyset},\quad \sigma\in \Gamma_n.
\end{equation}
Let $j\in\mathbb{N}_n$ and $\sigma\in\Gamma_n$. Then, by (\ref{eq-hat-ONB-vector}) and Proposition~\ref{pro-operator-eigenvector},
we find
\begin{equation*}
  \Xi_j^{(\nu)}\widehat{Z}_{\sigma}^{(\nu)}
  =\frac{1}{\sqrt{2^{n+1}}}\Xi_j^{(\nu)}\widehat{\Xi}_{\sigma}^{(\nu)}Z_{\emptyset}
  =\frac{1}{\sqrt{2^{n+1}}}\mathcal{E}_{\sigma}(j)\widehat{\Xi}_{\sigma}^{(\nu)}Z_{\emptyset}
  =\mathcal{E}_{\sigma}(j)\widehat{Z}_{\sigma}^{(\nu)}.
\end{equation*}
Thus (\ref{eq-hat-ONB}) holds. Next, let us show that $\widehat{\mathfrak{Z}}_n^{(\nu)}$ is an ONB for $\mathfrak{h}_n$.

Suppose $\sigma$, $\tau\in\Gamma_{n}$ with $\sigma\ne\tau$. Then, there is $j\in\mathbb{N}_{n}$ such that
$\mathcal{E}_{\sigma}(j)+\mathcal{E}_{\tau}(j)=0$ and $\mathcal{E}_{\sigma}(j)\mathcal{E}_{\tau}(j)=-1$, hence
\begin{equation*}
  \big(I_{\mathfrak{h}_n}+\mathcal{E}_{\sigma}(j)\Xi_j^{(\nu)}\big)\big(I_{\mathfrak{h}_n}+\mathcal{E}_{\tau}(j)\Xi_j^{(\nu)}\big)
  = I_{\mathfrak{h}_n}+\big(\mathcal{E}_{\sigma}(j)+\mathcal{E}_{\tau}(j)\big)\Xi_j^{(\nu)}
     + \mathcal{E}_{\sigma}(j)\mathcal{E}_{\tau}(j)\big(\Xi_j^{(\nu)}\big)^2
  = 0,
\end{equation*}
which together with (\ref{eq-hat-product-operator}) gives
\begin{equation*}
  \widehat{\Xi}_{\sigma}^{(\nu)}\widehat{\Xi}_{\tau}^{(\nu)}
   = \big[I_{\mathfrak{h}_n}+\mathcal{E}_{\sigma}(j)\Xi_j^{(\nu)}\big]
      \big[I_{\mathfrak{h}_n}+\mathcal{E}_{\tau}(j)\Xi_j^{(\nu)}\big]
      \prod_{k=0,k\ne j}^n\big(I_{\mathfrak{h}_n}+\mathcal{E}_{\sigma}(k)\Xi_k^{(\nu)}\big)
      \big(I_{\mathfrak{h}_n}+\mathcal{E}_{\tau}(k)\Xi_k^{(\nu)}\big)
   = 0.
\end{equation*}
Thus, by the self-adjoint property of $\widehat{\Xi}_{\sigma}^{(\nu)}$, we get
\begin{equation*}
\big\langle \widehat{Z}_{\sigma}^{(\nu)},\widehat{Z}_{\tau}^{(\nu)}\big\rangle_{\mathfrak{h}}
=\frac{1}{2^{n+1}}\big\langle \widehat{\Xi}_{\sigma}^{(\nu)}Z_{\emptyset},\widehat{\Xi}_{\tau}^{(\nu)}Z_{\emptyset}\big\rangle_{\mathfrak{h}}
=\frac{1}{2^{n+1}}\big\langle Z_{\emptyset},\widehat{\Xi}_{\sigma}^{(\nu)}\widehat{\Xi}_{\tau}^{(\nu)}Z_{\emptyset}\big\rangle_{\mathfrak{h}}
=0.
\end{equation*}
On the other hand, from (\ref{eq-hat-product-operator}) and by careful calculations, we find
\begin{equation*}
  \widehat{\Xi}_{\sigma}^{(\nu)} = \sum_{\gamma\in \Gamma_n}(-1)^{\#(\gamma\setminus \sigma)}\prod_{j\in \gamma}\Xi_j^{(\nu)},
\end{equation*}
where $(-1)^{\#(\gamma\setminus \sigma)}=1$ when $\gamma\setminus \sigma=\emptyset$, and by convention
$\prod_{j\in \gamma}\Xi_j^{(\nu)}=I_{\mathfrak{h}_n}$ when $\gamma=\emptyset$. This, together with (\ref{eq-hat-ONB-vector}) and Proposition~\ref{prop-shift-commutative}, gives
\begin{equation}\label{eq-MBV-formula}
  \widehat{Z}_{\sigma}^{(\nu)}
  = \frac{1}{\sqrt{2^{n+1}}}\sum_{\gamma\in \Gamma_n}(-1)^{\#(\gamma\setminus \sigma)}e^{-\mathrm{i}\sum_{j\in \gamma}\theta_j}Z_{\gamma},
\end{equation}
which further yields
\begin{equation*}
\big\langle \widehat{Z}_{\sigma}^{(\nu)},\widehat{Z}_{\sigma}^{(\nu)}\big\rangle_{\mathfrak{h}}
 = \frac{1}{2^{n+1}}\sum_{\gamma\in \Gamma_n}(-1)^{2\#(\gamma\setminus \sigma)}\big|e^{-\mathrm{i}\sum_{j\in \gamma}\nu_k}\big|^2
 = 1.
\end{equation*}
Thus $\widehat{\mathfrak{Z}}_n^{(\nu)}$ is an orthonormal system in $\mathfrak{h}_n$, which, together with
$\#(\widehat{\mathfrak{Z}}_n^{(\nu)})=\dim \mathfrak{h}_n$,
implies that $\widehat{\mathfrak{Z}}_n^{(\nu)}$ is an ONB for $\mathfrak{h}_n$.
\end{proof}

Theorem~\ref{thr-hat-ONB} makes one get an ONB
$\big\{\widehat{Z}_{\sigma}^{(\nu)}\otimes e_j\mid \sigma\in\Gamma_n,\, 0\leq j\leq d_{\mathcal{K}}-1\big\}$
for the tensor space $\mathfrak{h}_{n}\otimes \mathcal{K}$.
Hence each $\Psi\in\mathfrak{h}_n\otimes \mathcal{K}$ has an expansion of the form
\begin{equation}\label{eq-K-component}
\Psi=\sum_{\sigma\in\Gamma_n}\widehat{Z}_{\sigma}^{(\nu)}\otimes u_{\sigma}^{(\nu)},
\end{equation}
where $u_{\sigma}^{(\nu)}=\sum_{j=0}^{d_{\mathcal{K}}-1}\big\langle\widehat{Z}_{\sigma}^{(\nu)}\otimes e_j,\Psi\big\rangle e_j$,
which we call the $\sigma$th $\mathcal{K}$-component of $\Psi$ with respect to $\widehat{\mathfrak{Z}}_n^{(\nu)}$.

\begin{proposition}\label{prop-4-3}
Let $\{u_{\sigma} \mid \sigma\in \Gamma_n\}$ and $\{v_{\sigma} \mid \sigma\in \Gamma_n\}$ be two sets of vectors in
$\mathcal{K}$. Suppose that
\begin{equation*}
  \sum_{\sigma\in\Gamma_n}\widehat{Z}_{\sigma}^{(\nu)}\otimes u_{\sigma}
  = \sum_{\sigma\in\Gamma_n}\widehat{Z}_{\sigma}^{(\nu)}\otimes v_{\sigma}.
\end{equation*}
Then $u_{\sigma}= v_{\sigma}$, $\forall\,\sigma\in \Gamma_n$.
\end{proposition}

\begin{proof}
Let $\tau\in \Gamma_n$. Then, for all $j$ with $0\leq j \leq d_{\mathcal{K}}-1$, straightforward calculations give
\begin{equation*}
  \langle e_j, u_{\tau}\rangle_{\mathcal{K}}
  = \Big\langle \widehat{Z}_{\tau}^{(\nu)}\otimes e_j,\sum_{\sigma\in\Gamma_n}\widehat{Z}_{\sigma}^{(\nu)}\otimes u_{\sigma}\Big\rangle
  = \Big\langle \widehat{Z}_{\tau}^{(\nu)}\otimes e_j,\sum_{\sigma\in\Gamma_n}\widehat{Z}_{\sigma}^{(\nu)}\otimes v_{\sigma}\Big\rangle
  = \langle e_j, v_{\tau}\rangle_{\mathcal{K}},
\end{equation*}
which, together with the fact that $\{e_j \mid 0\leq j \leq d_{\mathcal{K}}-1\}$ is an ONB for $\mathcal{K}$, implies
$u_{\tau}=v_{\tau}$.
\end{proof}

Recall that, for $\sigma\in \Gamma_n$, the algebraic sum of the system $\mathfrak{C}$ corresponding to $\sigma$
is a unitary operator on $\mathcal{K}$. The next theorem then reveals a close relationship between such a unitary operator
and $\mathsf{W}^{(\nu)}$.

\begin{theorem}\label{thr-W-U-relation}
For all $\sigma\in\Gamma_n$ and $u\in \mathcal{K}$, it holds that
$\mathsf{W}^{(\nu)}\big(\widehat{Z}_{\sigma}^{(\nu)}\otimes u\big)
=\widehat{Z}_{\sigma}^{(\nu)}\otimes\big(U_{\sigma}u\big)$.
\end{theorem}

\begin{proof}
Let $\sigma\in\Gamma_n$ and $u\in \mathcal{K}$. With Theorem~\ref{thr-hat-ONB}
as well as the definitions of $\mathsf{W}^{(\nu)}$ and $U_{\sigma}$, we get
\begin{equation*}
\begin{aligned}
\mathsf{W}^{(\nu)}\big(\widehat{Z}_{\sigma}^{(\nu)}\otimes u\big)
& = \sum_{j=0}^n \big(\Xi_j^{(\nu)}\otimes C_j\big)\big(\widehat{Z}_{\sigma}^{(\nu)}\otimes u\big)\\
& = \sum_{j=0}^n \big(\mathcal{E}_{\sigma}(j)\widehat{Z}_{\sigma}^{(\nu)}\big)\otimes(C_ju)\\
& = \widehat{Z}_{\sigma}^{(\nu)}\otimes\Big(\sum_{j=0}^n\mathcal{E}_{\sigma}(j)C_j\Big)u\\
&=\widehat{Z}_{\sigma}^{(\nu)}\otimes\big(U_{\sigma}u\big),
\end{aligned}
\end{equation*}
which is the desired.
\end{proof}

We now turn to $\mathsf{W}^{(\nu)}$ and examine its spectral property. Recall that the point-spectrum of an operator
is the set consisting of all its eigenvalues. The next theorem describes close links
between the point-spectrum of $\mathsf{W}^{(\nu)}$ and those of operators $U_{\sigma}$, $\sigma\in \Gamma_n$,
which are completely determined by the coin operator system $\mathfrak{C}$.

\begin{theorem}\label{thr-point-spectrum}
Denote by $\mathrm{Spec}^{(p)} (T)$ the point-spectrum of an operator $T$. Then, it holds that
\begin{equation}\label{eq}
  \mathrm{Spec}^{(p)}(\mathsf{W}^{(\nu)})= \bigcup_{\sigma\in \Gamma_n}\mathrm{Spec}^{(p)}(U_{\sigma}).
\end{equation}
\end{theorem}

\begin{proof}
Suppose $\lambda$ belongs to the union of $\big\{\mathrm{Spec}^{(p)}(U_{\sigma}) \mid \sigma\in \Gamma_n\big\}$.
Then, there exists $\sigma\in \Gamma_n$ and $u\in \mathcal{K}$ with $u\ne 0$ such that $U_{\sigma}u=\lambda u$.
Apparently, $\widehat{Z}_{\sigma}^{(\nu)}\otimes u\ne 0$. On the other hand, by Theorem~\ref{thr-W-U-relation}, we have
\begin{equation*}
  \mathsf{W}^{(\nu)}\big(\widehat{Z}_{\sigma}^{(\nu)}\otimes u\big)
  =\widehat{Z}_{\sigma}^{(\nu)}\otimes\big(U_{\sigma}u\big)
  = \lambda \big(\widehat{Z}_{\sigma}^{(\nu)}\otimes u\big).
\end{equation*}
Thus, $\lambda$ is an eigenvalue of $\mathsf{W}^{(\nu)}$, namely $\lambda\in \mathrm{Spec}^{(p)}(\mathsf{W}^{(\nu)})$.

Conversely, suppose $\lambda\in \mathrm{Spec}^{(p)}(\mathsf{W}^{(\nu)})$. Then, there exists some
$\Psi\in \mathfrak{h}_n\otimes \mathcal{K}$ with $\Psi\ne 0$ such that $\mathsf{W}^{(\nu)}\Psi = \lambda\Psi$.
Let $\big\{u_{\sigma}^{(\nu)} \mid \sigma\in \Gamma_n\big\}$ be the $\mathcal{K}$-components of $\Psi$
with respect to $\widehat{\mathfrak{Z}}_n^{(\nu)}$ (see (\ref{eq-K-component}) and comments therein).
Then
\begin{equation*}
  \Psi=\sum_{\sigma\in\Gamma_n}\widehat{Z}_{\sigma}^{(\nu)}\otimes u_{\sigma}^{(\nu)}.
\end{equation*}
This, together with $\mathsf{W}^{(\nu)}\Psi = \lambda\Psi$ as well as Theorem~\ref{thr-W-U-relation}, yields
\begin{equation*}
  \sum_{\sigma\in\Gamma_n}\widehat{Z}_{\sigma}^{(\nu)}\otimes \big(U_{\sigma} u_{\sigma}^{(\nu)}\big)
  = \sum_{\sigma\in\Gamma_n}\widehat{Z}_{\sigma}^{(\nu)}\otimes \big(\lambda u_{\sigma}^{(\nu)}\big).
\end{equation*}
Thus, by Proposition~\ref{prop-4-3}, we have $U_{\sigma} u_{\sigma}^{(\nu)}= \lambda u_{\sigma}^{(\nu)}$, $\forall\, \sigma\in \Gamma_n$.
Noting that $\Psi\ne 0$, there exists some $\sigma\in \Gamma_n$ such that $u_{\sigma}^{(\nu)}\ne 0$, which
together with $U_{\sigma} u_{\sigma}^{(\nu)}= \lambda u_{\sigma}^{(\nu)}$ implies that $\lambda\in \mathrm{Spec}^{(p)}(U_{\sigma})$.
In particular, $\lambda$ belongs to the union of $\big\{\mathrm{Spec}^{(p)}(U_{\sigma}) \mid \sigma\in \Gamma_n\big\}$.
\end{proof}

\begin{definition}\label{def-approximate-spectrum}
Let $T$ be a bounded operator on a complex Hilbert space $\mathcal{H}$. A complex number $\lambda$ is called
an approximate eigenvalue of $T$ if there exists a sequence $(x_k)_{k\geq 1}$ of unit vectors in $\mathcal{H}$
such that $\|Tx_k- \lambda x_k\|_{\mathcal{H}}\rightarrow 0$ as $k\rightarrow\infty$.
The the approximate-spectrum of $T$, written $\mathrm{Aev}\,(T)$, is the set consisting of all approximate eigenvalue of $T$.
\end{definition}

By the definition, for a general bounded operator $T$ on a complex Hilbert space $\mathcal{H}$, it holds true that
\begin{equation}
\mathrm{Spec}^{(p)}\,(T)\subset \mathrm{Aev}\,(T)\subset \mathrm{Spec}\,(T).
\end{equation}
However, if $T$ is a self-adjoint bounded operator, then by the well-known Weyl's criterion (see, e.g, \cite{Borthwick}) one has
$\mathrm{Aev}\,(T)=\mathrm{Spec}\,(T)$,
namely $\mathrm{Aev}\,(T)$ coincides with the spectrum of $T$ in this case.
Thus, $\mathrm{Aev}\,(T)$ contains much information about the spectrum of $T$ and deserves to be studied.

\begin{theorem}\label{thr-approximate-spectrum}
It holds true that\ \
$\mathrm{Aev}\,(\mathsf{W}^{(\nu)})= \bigcup_{\sigma\in \Gamma_n}\mathrm{Aev}\,(U_{\sigma})$.
\end{theorem}

\begin{proof}
Let $\lambda$ be in the union of $\big\{\mathrm{Aev}\,(U_{\sigma})\mid \sigma\in \Gamma_n\big\}$.
Then, there exists some $\sigma\in \Gamma_n$ and a sequence $(u_k)_{k\geq 1}$ of unit vectors in $\mathcal{K}$
such that $\|U_{\sigma}u_k - \lambda u_k\|_{\mathcal{K}}\rightarrow 0$ as $k\rightarrow\infty$.
Clearly $\big(\widehat{Z}_{\sigma}^{(\nu)}\otimes u_k\big)_{k\geq 1}$ is a sequence of unit vectors in $\mathfrak{h}_n\otimes \mathcal{K}$.
On the other hand, using Theorem~\ref{thr-W-U-relation}, we find
\begin{equation*}
  \big\|\mathsf{W}^{(\nu)}\big(\widehat{Z}_{\sigma}^{(\nu)}\otimes u_k\big)- \lambda\big(\widehat{Z}_{\sigma}^{(\nu)}\otimes u_k\big)\big\|
  = \big\|\widehat{Z}_{\sigma}^{(\nu)}\otimes U_{\sigma}u_k - \widehat{Z}_{\sigma}^{(\nu)}\otimes (\lambda u_k)\big\|
  = \|U_{\sigma}u_k - \lambda u_k\|_{\mathcal{K}},\quad k\geq 1,
\end{equation*}
which implies that
$\big\|\mathsf{W}^{(\nu)}\big(\widehat{Z}_{\sigma}^{(\nu)}\otimes u_k\big)
 - \lambda\big(\widehat{Z}_{\sigma}^{(\nu)}\otimes u_k\big)\big\|\rightarrow 0$ as $k\rightarrow \infty$.
Thus $\lambda \in \mathrm{Aev}\,(\mathsf{W}^{(\nu)})$.

Conversely, let $\lambda \in \mathrm{Aev}\,(\mathsf{W}^{(\nu)})$. Then, there exists a sequence $(\Psi_k)_{k\geq 1}$ of unit vectors
in $\mathfrak{h}_n\otimes \mathcal{K}$ such that $\|\mathsf{W}^{(\nu)}\Psi_k - \lambda\Psi_k\|\rightarrow 0$ as $k\rightarrow\infty$.
For $\sigma\in \Gamma_n$ and $k\geq 1$, let $u_{\sigma}^{(\nu)}(k)$ be the $\sigma$th $\mathcal{K}$-component of $\Psi_k$
with respect to $\widehat{Z}_n^{(\nu)}$. Then
\begin{equation*}
  \Psi_k = \sum_{\sigma\in \Gamma_n} \widehat{Z}_{\sigma}^{(\nu)}\otimes u_{\sigma}^{(\nu)}(k),\quad k\geq 1,
\end{equation*}
which together with Theorem~\ref{thr-W-U-relation} implies that
\begin{equation*}
  \mathsf{W}^{(\nu)}\Psi_k = \sum_{\sigma\in \Gamma_n} \widehat{Z}_{\sigma}^{(\nu)}\otimes \big(U_{\sigma}u_{\sigma}^{(\nu)}(k)\big),\quad k\geq 1.
\end{equation*}
Thus, by a lengthy calculation, we get
\begin{equation*}
 \big\|\mathsf{W}^{(\nu)}\Psi_k -\lambda \Psi_k\big\|^2
 = \sum_{\sigma\in \Gamma_n} \big\|U_{\sigma}u_{\sigma}^{(\nu)}(k)- \lambda u_{\sigma}^{(\nu)}(k)\big\|_{\mathcal{K}}^2,\quad k\geq 1.
\end{equation*}
This, together with $\|\mathsf{W}^{(\nu)}\Psi_k - \lambda\Psi_k\|\rightarrow 0$, implies that
$\big\|U_{\sigma}u_{\sigma}^{(\nu)}(k)- \lambda u_{\sigma}^{(\nu)}(k)\big\|_{\mathcal{K}}\rightarrow 0$ for all $\sigma\in \Gamma_n$.
On the other hand, straightforward calculations yield
\begin{equation*}
  \sum_{\sigma\in \Gamma_n} \big\|u_{\sigma}^{(\nu)}(k)\big\|_{\mathcal{K}}^2= \|\Psi_k \|^2=1,\quad \forall\, k\geq 1.
\end{equation*}
This implies that there exists some $\sigma_0\in \Gamma_n$ such that $\big\|u_{\sigma_0}^{(\nu)}(k)\big\|_{\mathcal{K}}\nrightarrow 0$
as $k\rightarrow \infty$. Thus, there exists $\epsilon_0>0$ and a subsequence $\big(u_{\sigma_0}^{(\nu)}(k_j)\big)_{j\geq 1}$
such that $\|u_{\sigma_0}^{(\nu)}(k_j)\|_{\mathcal{K}}\geq \epsilon_0$, $\forall\, j\geq 1$. Now, by setting
\begin{equation*}
v_j= \frac{u_{\sigma_0}^{(\nu)}(k_j)}{\|u_{\sigma_0}^{(\nu)}(k_j)\|_{\mathcal{K}}},\quad j\geq 1,
\end{equation*}
we get a sequence $(v_j)_{j\geq 1}$ of unit vectors in $\mathcal{K}$, which satisfies that
\begin{equation*}
  \big\|U_{\sigma_0}v_j- \lambda v_j\big\|_{\mathcal{K}}
  = \frac{1}{\|u_{\sigma_0}^{(\nu)}(k_j)\|_{\mathcal{K}}}\big\|U_{\sigma_0}u_{\sigma_0}^{(\nu)}(k_j)
     - \lambda u_{\sigma_0}^{(\nu)}(k_j)\big\|_{\mathcal{K}}
  \leq \frac{1}{\epsilon_0}\big\|U_{\sigma_0}u_{\sigma_0}^{(\nu)}(k_j)
     - \lambda u_{\sigma_0}^{(\nu)}(k_j)\big\|_{\mathcal{K}},
\end{equation*}
where $j\geq 1$. Noting that
$\big\|U_{\sigma_0}u_{\sigma_0}^{(\nu)}(k_j)- \lambda u_{\sigma_0}^{(\nu)}(k_j)\big\|_{\mathcal{K}}\rightarrow 0$
as $j\rightarrow \infty$, we have
\begin{equation*}
\big\|U_{\sigma_0}v_j- \lambda v_j\big\|_{\mathcal{K}}\rightarrow 0\quad \mbox{as $j\rightarrow \infty$}.
\end{equation*}
Thus, $\lambda$ is an approximate eigenvalue of $U_{\sigma_0}$, namely $\lambda\in \mathrm{Aev}\,(U_{\sigma_0})$.
In particular, $\lambda$ belongs to the union of $\big\{\mathrm{Aev}\,(U_{\sigma})\mid \sigma\in \Gamma_n\big\}$.
\end{proof}

Apparently, the algebraic sums $\{U_{\sigma}\mid \sigma\in \Gamma_n\}$ are completely determined
by the coin operator system $\mathfrak{C}$. Hence, Theorem~\ref{thr-point-spectrum} and Theorem~\ref{thr-approximate-spectrum}
essentially give representations of $\mathrm{Spec}^{(p)}\,(W^{(\nu)})$ and $\mathrm{Aev}\,(W^{(\nu)})$ in terms of
the coin operator system $\mathfrak{C}$.

As is seen, the walk $W^{(\nu)}$ is dependent of the magnetic potential $\nu$ since so are its shift operators
$\big\{\Xi_j^{(\nu)} \mid j\in \mathbb{N}_n\big\}$. However, the next theorem shows that the walk $W^{(\nu)}$ has spectral stability
with respect to the magnetic potential $\nu$.

\begin{theorem}\label{thr-spectral-stability}
Both $\mathrm{Spec}^{(p)}\,(W^{(\nu)})$ and $\mathrm{Aev}\,(W^{(\nu)})$ are completely independent of the magnetic potential $\nu$.
\end{theorem}

\begin{proof}
From construction (\ref{eq-weighted-coin-sum}), we easily see that the algebraic sums $\big\{U_{\sigma} \mid \sigma\in\Gamma_n\big\}$
of the coin operator system $\mathfrak{C}$ are completely independent of the magnetic potential $\nu$
since so is the system $\mathfrak{C}$ itself. This, together with Theorems~\ref{thr-point-spectrum} and \ref{thr-approximate-spectrum},
implies that $\mathrm{Spec}^{(p)}\,(W^{(\nu)})$ and $\mathrm{Aev}\,(W^{(\nu)})$ are completely independent of the magnetic potential $\nu$.
\end{proof}

\section{Conclusion remarks}

As is seen, in this paper we introduce a model of magnetic quantum walk actually on a general hypercube
by using quantum Bernoulli noises, and investigate its behavior from a viewpoint of spectral theory.
As mentioned above, in \cite{W-JMP-2022} the author introduced a quantum walk in the same setting as in the present paper,
and considered its behavior from a perspective of probability distribution.
Comparing the walk in this paper with that in \cite{W-JMP-2022}, we come to some observations bellow.

For the walk introduced in \cite{W-JMP-2022}, its evolution operator just took the following form
\begin{equation}\label{eq}
\mathsf{W}\!_{\mathfrak{C}}= \sum_{k=0}^n (\partial_k^* + \partial_k)\otimes C_k,
\end{equation}
where $\mathfrak{C}=\{C_k \mid k\in \mathbb{N}_n\}$ is the coin operator system.
Clearly, $\mathsf{W}\!_{\mathfrak{C}}=\mathsf{W}^{(\nu)}$ with $\nu$ being the null magnetic potential.
In other words, the walk introduced in \cite{W-JMP-2022} is a very special case of the magnetic quantum walk
introduced in this paper. Therefore, as the main results in this paper, Theorem~\ref{thr-point-spectrum} and
Theorem~\ref{thr-approximate-spectrum} are definitely effective when applied to the walk in \cite{W-JMP-2022}.

As can be seen, the shift operators $\partial_k^* + \partial_k$ in \cite{W-JMP-2022} are a special case of
the magnetic shift operators $\Xi_k^{(\nu)} = e^{-\mathrm{i}\nu_k}\partial_k^* + e^{\mathrm{i}\nu_k}\partial_k$ in this paper.
Thus, from a mathematical perspective, the technical propositions and theorems (e.g. Theorems~\ref{thr-3-1}, \ref{thr-evolution-operator},
\ref{thr-hat-ONB} and \ref{thr-W-U-relation}) proven in this paper generalize those in \cite{W-JMP-2022}.

Finally, the magnetic quantum walk introduced in this paper can also be analyzed from a viewpoint of probability distribution
and similar results can be obtained with the same arguments as in \cite{W-JMP-2022}.

\section*{Author declarations}

\subsection*{Conflict of interest}

The author has no conflicts to disclose.

\subsection*{Data Availability}

Data sharing is not applicable to this article as no new data were created or analyzed in this study.

\end{document}